\documentstyle[epsfig,11pt]{article}
\def\fr{\frac}
\def\t{\theta}

\newcommand{\be}{\begin{equation}}
\newcommand{\ee}{\end{equation}}
\newcommand{\ben}{\begin{eqnarray}}
\newcommand{\een}{\end{eqnarray}}
\newcommand{\bc}{\begin{center}}
\newcommand{\ec}{\end{center}}

\begin{document}

\title{Macrolensing signatures of large-scale violations of the weak
energy condition}

\author{Margarita Safonova$^{a}$\thanks{rita@iucaa.ernet.in},
Diego F. Torres$^{b}$\thanks{dtorres@venus.fisica.unlp.edu.ar} and
Gustavo E. Romero$^{b}$\thanks{romero@irma.iar.unlp.edu.ar}\\
{\small $^a$Department of Physics and Astrophysics, University of Delhi,
New Delhi--7, India} \\ {\small $^b$Instituto Argentino de
Radioastronom\'{\i}a, C.C.5, 1894 Villa Elisa, Buenos Aires,
Argentina} }

\date{\today }

\maketitle

\begin{abstract}
We present a set of simulations of the macrolensing effects
produced by large-scale cosmological violations of the energy
conditions. These simulations show how the appearance of a
background field of galaxies is affected when lensed by a region
with an energy density equivalent to a negative mass ranging from
$10^{12}$ to $10^{17}$ $|M_\odot|$. We compare with the
macrolensing results of equal amounts of positive mass, and show
that, contrary to the usual case where tangential arc-like
structures are expected, there appear radial arcs---runaway
filaments---and a central void. These results make the cosmological
macrolensing produced by space-time domains where the weak energy
conditions is violated, observationally distinguishable from
standard regions. Whether large domains with negative energy
density indeed exist in the universe can now be decided by future
observations of deep fields.
\end{abstract}



\newpage

\subsection*{Introduction}

The energy conditions (EC) of classical General Relativity (see
Appendix) are conjectures widely used to prove theorems concerning
singularities and black hole thermodynamics. The area increase
theorem for black holes, the topological censorship theorem, and
the singularity theorem of stellar collapse, are among the most
important ones \cite{VISSER-BOOK}. However, the set of EC
constitute only plausible statements, all of them lacking a
rigorous proof. Moreover, several situations in which the EC are
violated are known; perhaps the most quoted being the Casimir
effect. For other physical situations see Ref. \cite{nec} and
references therein. Typically, observed violations are produced by
small quantum systems, resulting of the order of $\hbar$. It is
currently far from clear whether there could be macroscopic
quantities of such an exotic, EC-violating, matter.\\

The violations of the EC, in particular the weak one, would admit
the existence of negative amounts of mass. As Bondi remarked in
Ref. \cite{BONDI}, it is an empirical fact that inertial and
gravitational masses are both positive quantities. In fact, the
possible existence of negative gravitational masses is being
investigated at least since the end of the nineteen century
\cite{JAMMER}. The empirical absence of negative masses in the
Earth neighborhood could be explained as the result of the
plausible assumption that, repelled by the positive masses
prevalent in our region of space, the negative ones have been
driven away to extragalactic distances. Recently, primordially
formed negative gravitational masses have been proposed as an
explanation of the voids observed in the extragalactic space
\cite{PIRAN-0}. Mann \cite{MANN} have found, in addition, that
dense regions of negative mass can undergo gravitational collapse,
probably ending up in exotic black holes.\\

Of all the systems which would require violations of the EC in
order to exist, wormholes are the most intriguing
\cite{wormhole-papers}. The most salient feature of these objects
is that an embedding of one of their space-like sections in
Euclidean space displays two asymptotically flat regions joined by
a throat. Observational properties of stellar and sub-stellar size
wormholes have been recently discussed in the literature; the
propagation of perturbations and particles \cite{karsan}, the
possible violation of the equivalence principle \cite{e}, the
gravitational lensing of light \cite{cramer}, and the possibility
of relating them with gamma-ray bursts \cite{diego-grwh} are some
examples. In all these cases, however, wormholes were considered
as compact objects formed by about 1 negative solar mass.\\

Very recently, the consequences of the validity of the EC were
confronted with possible values of the Hubble parameter and the
gravitational redshifts of the oldest stars \cite{VISSER-HUBBLE}.
It was deduced that the strong energy condition (considering the
density of the universe as a whole) can be violated rather late in
the history of the universe, sometime between the formation of the
oldest stars and the present epoch. This would imply the existence
of a massive scalar field or a positive cosmological constant,
something what recent experiments seem to favor. But even if the
global energy density of the universe is WEC-respecting, we can
wonder whether there exist space-time domains where large-scale
violations of the EC occur, allowing the formation of physical
systems with an energy density equivalent to a total negative mass
of the size of a galaxy or even a cluster of galaxies.\\

If the answer to this question is yes, it is clear that we should
be able to see some gravitational effects on the light that
traverse such regions. In particular, lensing of distant background
sources should exhibit distinctive features. In what follows, we
present the results of a set of simulations showing the
macrolensing effects we could observe if such a large amount of
negative energy density exist in our universe.

\subsection*{Lens equation and description of the algorithm}

We begin the discussion of gravitational lensing by defining two
convenient planes, i.e. the source and lens plane. These planes,
described by Cartesian coordinate systems ($\beta_1,\beta_2$) and
($\t_1,\t_2$), pass through the source and the deflecting mass,
respectively, and are perpendicular to the optical axis (the
straight line joining the source plane, through the deflecting
mass, and the observer). We write the gravitational lens
equation, which governs the mapping from the lens to the source
plane, in dimensionless form
\be 
\mbox{\boldmath $\beta$} = \mbox{\boldmath $\theta$} 
\left [1 + \frac{\theta^2_{\rm E}}{\theta^2} \right]\,\,, \label{eq:leq}
\ee 
where {\boldmath$\beta$}({\boldmath$\theta$}) refers to 
the position of the
source (image) on the sky, as described in the Fig.~1. We have
defined the quantity $\theta_{\rm E}$ in analogy with lensing by
positive matter as
\be 
\theta^2_{\rm E} = \frac{ 4 G |M|}{c^2} \frac{D_{\rm ls}}
{D_{\rm s} D_{\rm l}} \,\,,
\ee 
where $D_{\rm l}$, $D_{\rm s}$, and $D_{\rm ls}$ are the angular
diameter distances from the observer to the lens, from the
observer to the source, and from the lens to the source,
respectively. We forward the reader looking for a complete
theoretical treatment of lensing by negative masses to
Ref.~\cite{US}. It should be borne in mind that although
$\theta_{\rm E}$ is the only natural angular scale in Eq.
(\ref{eq:leq}), it does not have the same physical meaning as in
the case of the positive lensing, since negative matter cannot
produce an Einstein ring.\\

\begin{figure}[t]
\centerline{
\includegraphics[width=0.3\textwidth]{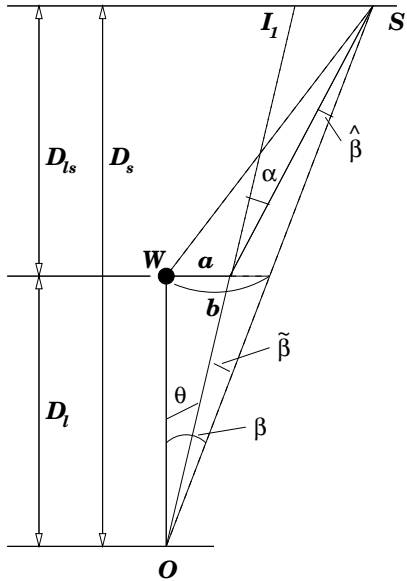}}
\caption{Lensing geometry for a negative mass. $O$ is the
observer, $S$ is the source, $W$ is the position of the negative
mass and $I_1$ is one the images. $\beta$ is the angle between the
source and the image, $\alpha$ is the deflection angle. $b$ is the
impact parameter, and all other quantities are auxiliary to obtain
the lens equation. $D_{\rm l}$, $D_{\rm s}$ and $D_{\rm ls}$ are angular
diameter distances.}
\end{figure}

\mbox{}For the simulations that follow, we have used a background
cosmology described by a Friedmann-Robertson-Walker flat universe
with $\Omega_{\rm m} = 1$ and a zero cosmological constant. In all
numerical computations we have used a Hubble parameter $h$ equal
to 1 ($H_0 = 100\,h$ km sec$^{-1}$ Mpc$^{-1}$). The relationship
$D_{\rm eff}=D_{\rm l} D_{\rm ls} D_{\rm s}^{-1}$, entering the Eq.~2,
is a measure of
the lensing efficiency of a given mass distribution. $D_{\rm eff}$
peaks, quite independently of the cosmological model assumed, at a
lens redshift of $\sim 0.2-0.4$ for sources at a typical redshift
$z_{\rm s} \sim 1-1.3$ \cite{3}. To be more realistic, we shall
place the lens at a redshift of $z_l = 0.3$ and  generate a random
sample of galaxies in the redshift range $ 0.3 < z_{\rm source} <
2.0$. The redshift distribution conserves their comoving number
density. This number density and the projected sizes for these
background sources were taken to be close to the Tyson population
of faint blue galaxies \cite{5}. The luminous area of each galaxy
was taken to be a circular disk of radius $R$ with a uniform
brightness profile and orientations of disk galaxies were randomly 
placed in space. {\it NOTE: the next lines you can remove if you want or 
retain}. This was done by defining in the code the ellipticity $e$ 
($e = (1-r)/(1+r)$, $r$ is the ratio of the minor axis to the major axis) 
and the position angle $\varphi$, and randomly chosing the values of
$e$ from the range $0<e<0.7$ and the values of $\varphi$ from
the range $0<\varphi<2\pi$. \footnote{The random number
generator needed in the code was taken from the book by Press et
al. \cite{1}, and we use the algorithm described in
Ref. \cite{2}, p.298. PGPLOT routine PGGRAY was employed in the code.}\\

The lens equation (\ref{eq:leq}) describes a mapping ${\bf \t}
\mapsto {\bf \beta}$, from the lens to the source plane. For
convenience, we redefine the lens plane as ${\bf x}$ and the
source plane as ${\bf y}$. Then, Eq. (\ref{eq:leq}) can be written
as 
\be
{\bf y} = {\bf x} \left(1+ \fr{\t^2_{\rm E}}{x^2}\right), 
\ee
where $x=|{\bf x}|=\sqrt{x_1^2+x_2^2}$.

We now consider a source, whose shape---either circular or
elliptical---can be described by a function $\chi({\bf y})$. Curves
of constant $\chi$ are the contours of the source. One can as well
consider $\chi$ as a function of ${\bf x}$, $\chi({\bf y}({\bf
x}))$, where ${\bf y}({\bf x})$ is found using the lens equation.
Thus, all points ${\bf x}$ of constant $\chi$ are mapped onto
points ${\bf y}$, which have a distance $\sqrt{\chi}$ from the
centre of the source. If the latter contour can be considered an
isophote of a source, one has thus found the corresponding
isophotes of the images. See Ref. \cite{2} for further details.

\subsection*{Simulations results}

In the first set of figures (Figs. 2-6), we show the results of
our simulations. Some special precautions must be taken for the 
largest masses. The
problem is that for a very massive lens the Einstein ring becomes
very large. Since for the negative mass lensing all sources inside
the double Einstein radius are shadowed (i.e. we can see the
images of only those sources which are outside the double Einstein
radius), if we were to use for lensing only the sources shown on
the window, a fold-four symmetry pattern would appear (Fig.~7). Only the
sources at the corners of the current window (and outside the
double Einstein radius) are lensed. In order to solve this
problem we have to consider also the sources from outside the
current window; then the lensing picture is restored and the scale
of the simulation is consistently increased. For this reason we
increase the number of background galaxies in Figs.~4--6. We show
this in detail in Figs.~7 and 8.\\

\begin{figure}
\centerline{
\includegraphics[width=0.8\textwidth]{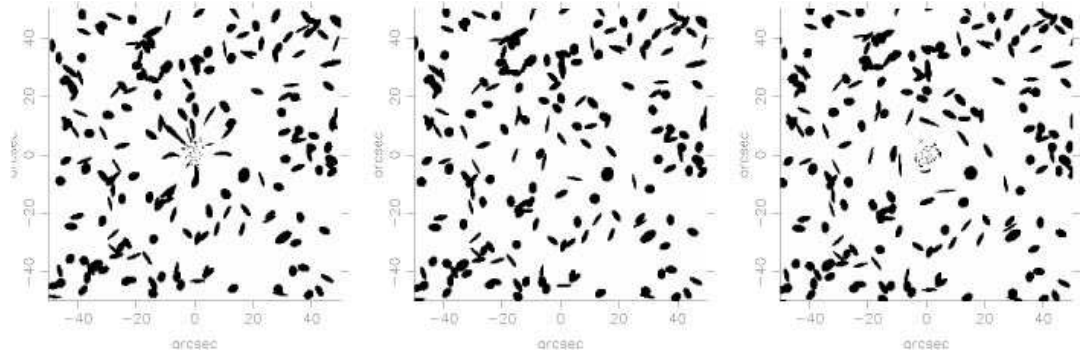}}
\caption{Left: Appearance of a background field of sources in a
range of redshifts (see text) (200 galaxies, intrinsic radius 7
Kpc), when lensed by a negative mass of $|M|_{\rm lens} = 1\times
10^{13}\,M_{\odot}$. Center: Unlensed background field. Right:
Appearance of the same background field of galaxies when it is
lensed by an equal amount of positive mass; redhsifts are the same
as for the negative mass case.}
\end{figure}

\begin{figure}
\centerline{
\includegraphics[width=0.8\textwidth]{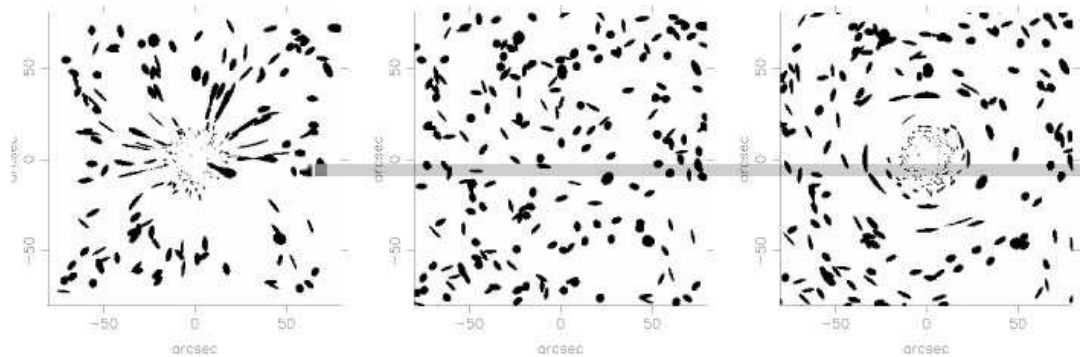}}
\caption{Left: Appearance of a background field of sources (200
galaxies, intrinsic radius 10 Kpc), when it is lensed by a
negative mass of $|M|_{\rm lens} = 1\times 10^{14}\,M_{\odot}$.
Center: As in Figure 2. Right: As in Figure 2.}
\end{figure}

\begin{figure}
\centerline{
\includegraphics[width=0.8\textwidth]{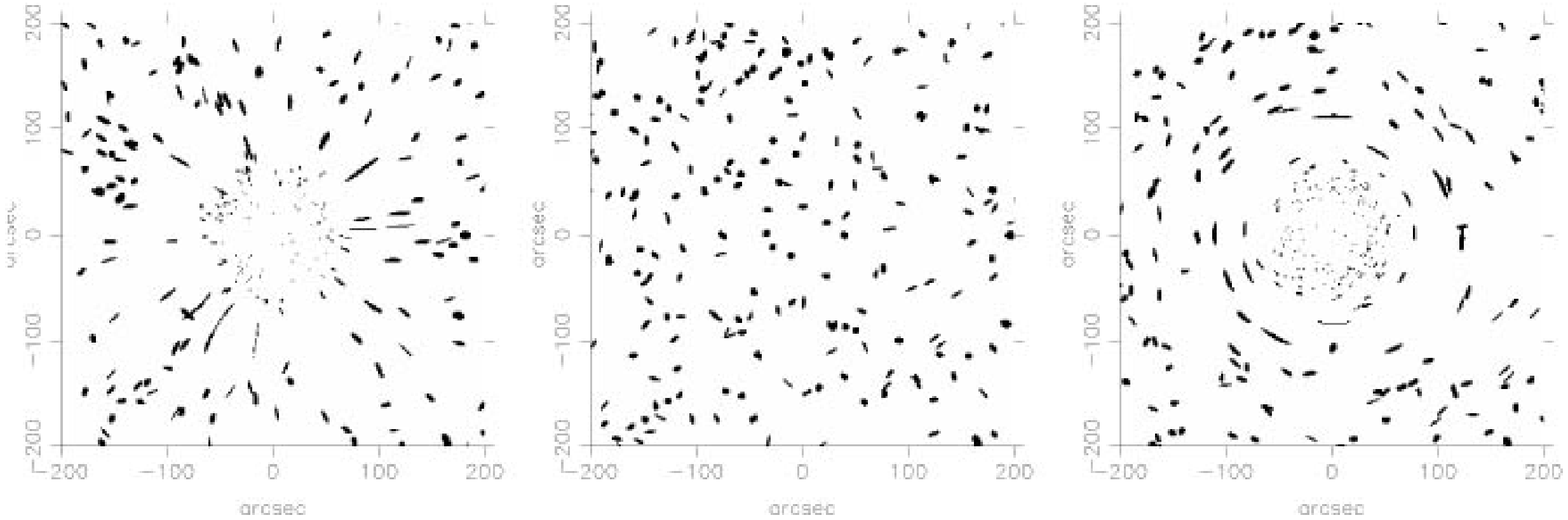}}
\caption{Left: Appearance of a background field of sources (300
galaxies, intrinsic radius 15 Kpc), when it is lensed by a
negative mass of $|M|_{\rm lens} = 1\times 10^{15}\,M_{\odot}$.
The simulation was made taking into account sources located within
1.2 of the size of the shown window. Center: As in Figure 2.
Right: As in Figure 2.}
\end{figure}

\begin{figure}
\centerline{
\includegraphics[width=0.8\textwidth]{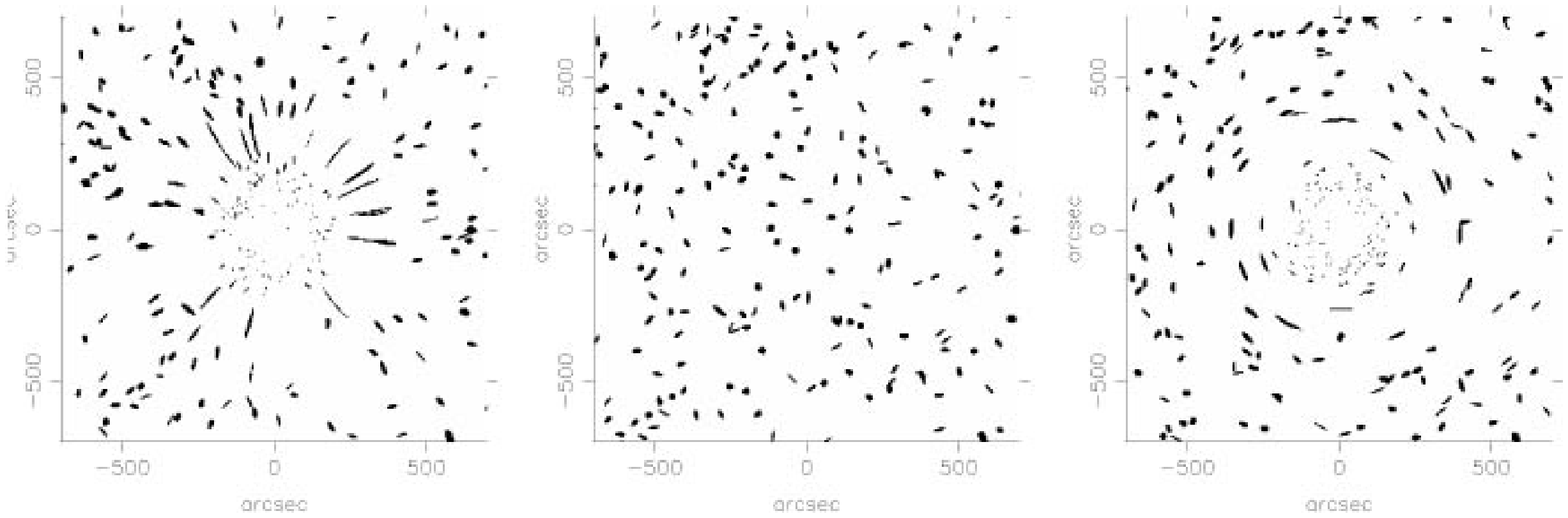}}
\caption{Left: Appearance of a background field of sources (300
galaxies, intrinsic radius 50 Kpc), when it is lensed by a
negative mass of $|M|_{\rm lens} = 1\times 10^{16}\,M_{\odot}$.
The simulation was made taking into account sources located within
1.2 of the size of the shown window. Center: As in Figure 2.
Right: As in Figure 2.}
\end{figure}

\begin{figure}
\centerline{
\includegraphics[width=0.8\textwidth]{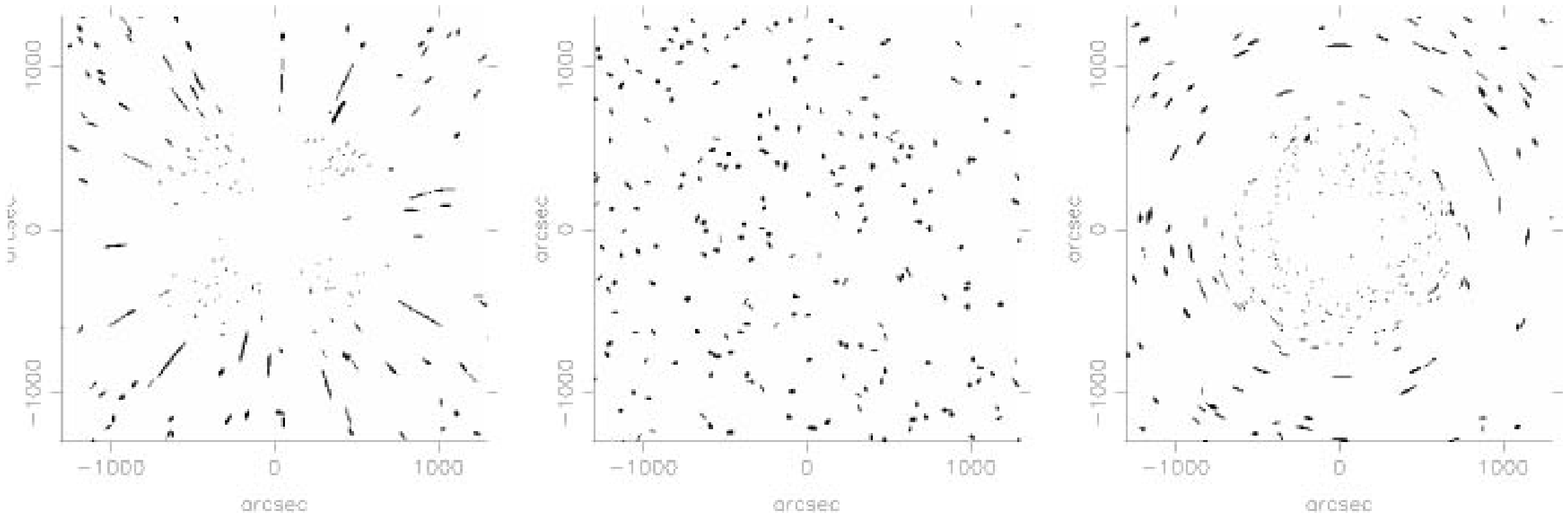}}
\caption{Left: Appearance of a background field of sources (500
galaxies), intrinsic radius 60 Kpc, when it is lensed by a
negative mass of $|M|_{\rm lens} = 1\times 10^{17}\,M_{\odot}$.
The simulation was actually made taking into account sources
located within 1.5 of the size of the shown window. Center: As in
Figure 2. Right: As in Figure 2.}
\end{figure}

\begin{figure}
\centerline{
\includegraphics[width=0.8\textwidth]{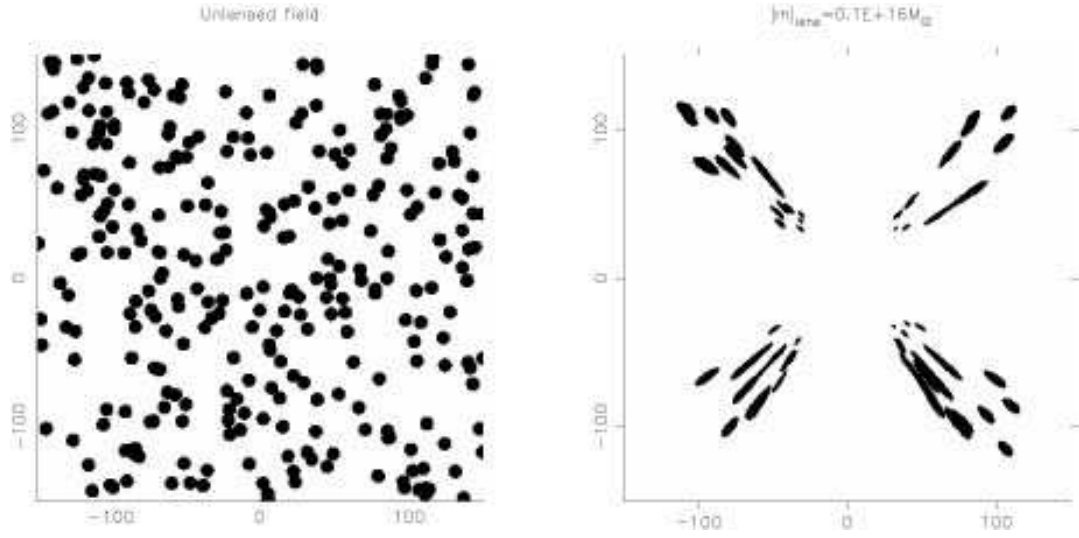}}
\caption{The need for the increase of the background number of
sources is shown by the appearance of a four-fold symmetric
pattern, which occurs because only the galaxies at the corners of
the left window are being affected by lensing effects. Axis are
marked in arcseconds. See next figure.} 
\label{pr}
\end{figure}

\begin{figure}
\centerline{
\includegraphics[width=0.8\textwidth]{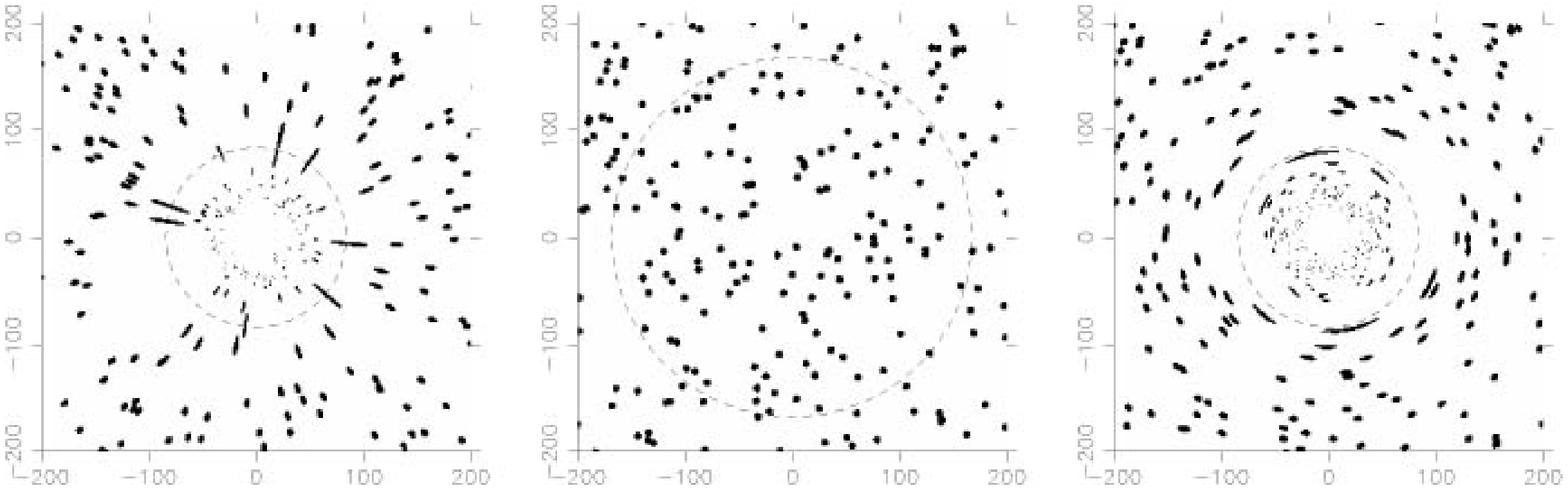}}
\caption{Left: Appearance of a background field of circular
sources (300 galaxies), each of them of 15 Kpc radius, when it is
lensed by a negative mass of $|M|_{\rm lens} = 1\times
10^{15}\,M_{\odot}$ with an Einstein angular radius equal to
$\theta_{\rm E}=84$ arcsec. The dashed circle is the Einstein
ring, the radial arcs are centered on it while the images inside
it are demagnified. Here $z_{\rm lens} = 0.4$, $z_{\rm
source}=1.4$. The simulation was made taking into account sources
located within 1.2 of the size of the shown window. Center:
Background field in the absence of the lens, dashed circle is the
double Einstein radius, all sources inside this radius are
shadowed. Right: Macrolensing effects produced by a positive mass
lens of $10^{15}\,M_{\odot}$, a dashed circle is the Einstein
radius shown here for comparison, sources inside it are strongly
lensed.} \label{rr}
\end{figure}

As a general feature of our simulations we can remark that,
opposite to the standard positive mass case, where ring-like
structures appear, the negative mass lensing produce finger-like,
apparently ``runaway" structures, which seem to escape from a
central void. This is in agreement with the appearance of a
central umbra in the case of a point-like negative mass lensing,
as studied by Cramer et al. \cite{cramer}. In the case of
macrolensing, the umbra (central void) is maintained on a larger
scale which, depending on the negative mass of the lens, can reach
hundreds of acrsec in linear size. This umbra is always larger
than the corresponding one generated in positive macrolensing (see
Figures 2-6) and totally different in nature \cite{cramer}. Then,
the existence of a macroscopic amount of negative mass lens can---
at least qualitatively---mimic the appearance of galaxy voids.

In order to explore the influence of the adopted redshift values,
we turn now to the case where $z_{\rm sources} = 0.08$ and 
$z_{\rm lens} = 0.05$. The Bootes void is the closest void to us, 
and lies between the supercluster Corona Borealis ($z \approx 
0.08$) and Hercules ($z \approx 0.03$) \cite{4}. 
This serves as motivation
for the selection of these redshift values. As an example of the
results for different lens masses, we show in Figs.~9 and~10 the
cases with $|M|_{\rm lens} = 1\times 10^{14}\,M_{\odot}$ and
$|M|_{\rm lens} = 1\times 10^{16}\,M_{\odot}$.
 
\begin{figure}
\centerline{
\includegraphics[width=0.8\textwidth]{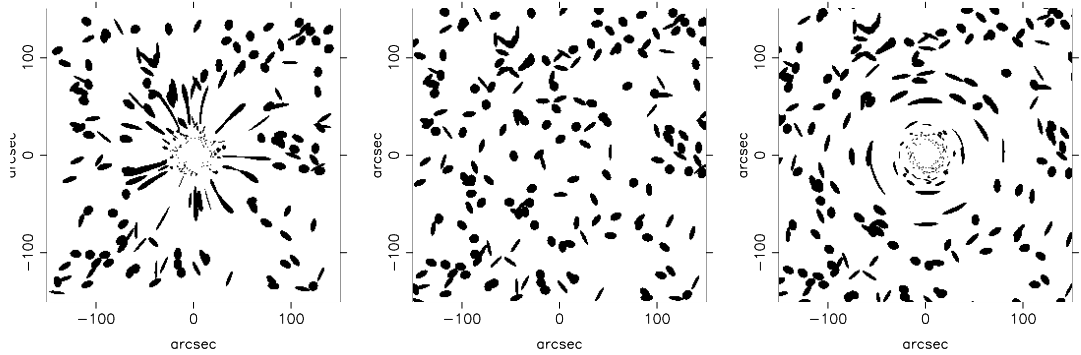}}
\caption{Left: Appearance of a background field of sources (200
galaxies), each of them 5 Kpc radius, when it is
lensed by a negative mass of $|M|_{\rm lens} = 1 \times 
10^{14}\,M_{\odot}$ with an Einstein angular radius equal to
$\theta_{\rm E}=47$ arcsec. Center: Unlensed background field.
Right: Appearance of the same background field of galaxies when
lensed by an equal amount of positive mass, located at the same
redshift (see text).} 
\end{figure}

\begin{figure}
\centerline{\epsfig{figure=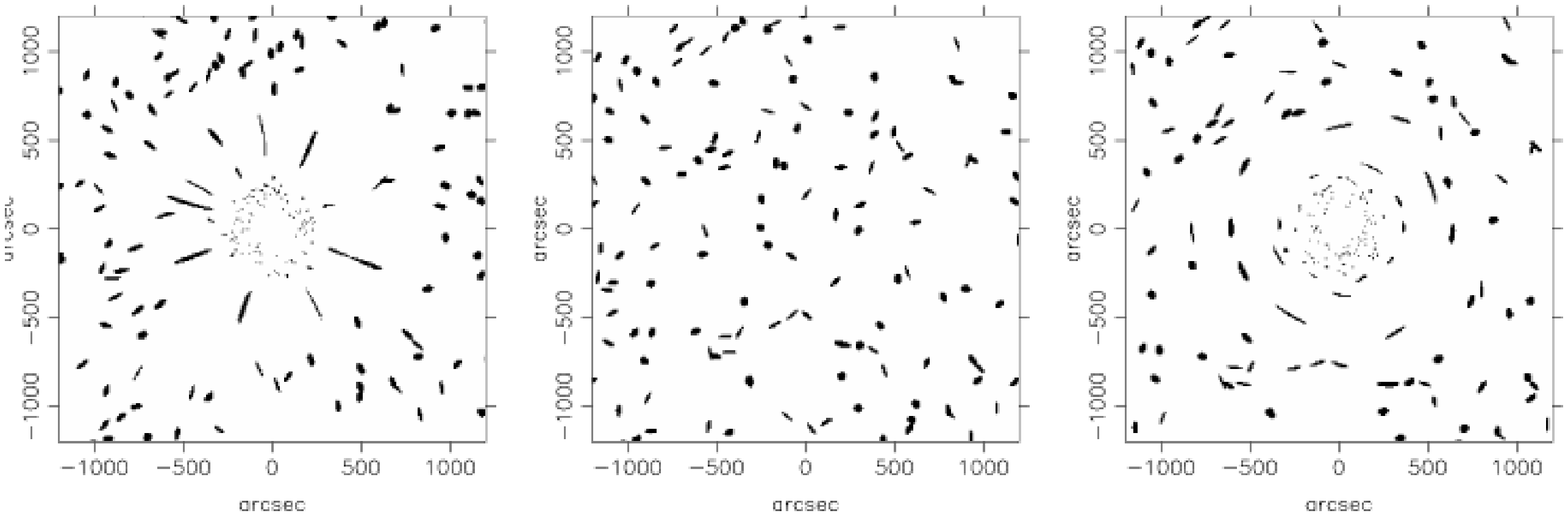, width=0.8\textwidth}}
\caption{Left: Appearance of a background field of sources (300
galaxies), each of them of 25 Kpc radius, when it is
lensed by a negative mass of $|M|_{\rm lens} = 1\times
10^{16}\,M_{\odot}$ with an Einstein angular radius equal to
$\theta_{\rm E}=467$ arcsec. Center: As in Figure 9. Right: As in
Figure 9.}
\end{figure}

It is interesting to note that although the background population
of galaxies is very dense, and for the standard model of cosmology
to be valid, one would expect a lot of lensing. However, there is still 
a surprising dearth of candidates for (positive mass) lensed sources
\cite{6}. Some of the richest clusters do not display arcs in the
deepest images. CL0016+16 ($z=0.56$), for instance, is one of the
richest and strongest X-ray emitter clusters. It is rather
extended on the sky, so the light from many background sources
should cross this cluster. Neither arcs nor arclets have been
found, though weak lensing has been reported. This may be pointing
towards a cautionary note: if the kind of finger-like structures
displayed in our figures is not directly seen in its full pattern,
that does not necessarily mean that they are absent. Even the
presence of one radial arc (without tangential counter arc and/or
tangential arcs) may be significant.

\subsection*{Concluding remarks}

The null EC (NEC) is the weakest of the EC. Usually, it was
considered that all reasonable forms of matter should at least
satisfy the NEC. However, even the NEC and its averaged version
(ANEC) are violated by quantum effects and semi-classical quantum
gravity (quantized matter fields in a classical gravitational
background). Moreover, it has recently been shown that there are
also large classical violations of the energy conditions
\cite{ULT}. This Letter shows that---disregarding the fundamental
mechanism by which the EC are violated, e.g. fundamental scalar
fields, modified gravitational theories, etc.---if large localized
violations of NEC exist in our universe, we can be able to detect
them through cosmological macrolensing. Contrary to the usual
case, where ring structures are expected, finger-like, ``runaway"
filaments and a central void appear. Figures 2--6 and 9--10 compare
with the case of macrolensing effects on background fields
produced by equal amounts of positive mass located at the same
redshift. Differences are notorious. These results make the
cosmological macrolensing produced by matter violating the weak
energy condition observationally distinguishable from the standard
situation. Whether large-scale violations of the EC resulting in
space-time regions with average negative energy density indeed
exist in the universe can now be decided through observations.

\subsection*{Acknowledgments}
MS is supported by a ICCR scholarship (Indo-Russian Exchange
programme) and acknowledges the hospitality of IUCAA, Pune. We
would like to deeply thank Tarun Deep Saini for his invaluable
help with the software. This work has also been supported by
CONICET (PIP 0430/98, GER), ANPCT (PICT 98 No. 03-04881, GER) and
Fundaci\'on Antorchas (through separate grants to DFT and GER).

\subsection*{Appendix: Energy conditions}

To specify what we are referring to when talking about the energy
conditions, we provide their point-wise form. They are the null
(NEC), the weak (WEC), the strong (SEC), and the dominant (DEC)
energy conditions. Using a Friedmann-Robertson-Walker space-time
and the Einstein field equations, they read:

\begin{eqnarray}
\hbox{NEC} & \iff &  \quad (\rho + p \geq 0 ), \nonumber \\
\hbox{WEC} & \iff & \quad (\rho \geq 0 ) \hbox{ and } (\rho + p
\geq 0),  \nonumber \\ \hbox{SEC} & \iff & \quad (\rho + 3 p \geq
0 ) \hbox{ and } (\rho + p \geq 0), \nonumber \\ \hbox{DEC} & \iff
& \quad (\rho \geq 0 ) \hbox{ and } (\rho \pm p \geq 0).
\end{eqnarray}

The EC are, then, linear relations between the energy and the
pressure of the matter generating the space-time curvature.

\end{document}